\documentclass{aa}

\usepackage{graphicx}
\usepackage{txfonts}
\usepackage{natbib}
\usepackage{xspace}
\usepackage[colorlinks=true,linkcolor=blue,citecolor=blue]{hyperref}


\newcommand{\hlink}[1]{\url{http://#1}\xspace}

\newcommand{\rfig}[1]{Fig.~\ref{#1}}

\newcommand{\rsec}[1]{Section \ref{#1}}
\newcommand{\rsecs}[1]{Sections \ref{#1}}

\newcommand{\herschel}{{\it Herschel}\xspace}
\newcommand{\spitzer}{{\it Spitzer}\xspace}
\newcommand{\hubble}{{\it Hubble}\xspace}
\newcommand{\hst}{{\it HST}\xspace}

\newcommand{\um}{\mu{\rm m}}

\newcommand{\uJy}{\mu{\rm Jy}}
\newcommand{\mJy}{{\rm mJy}}

\newcommand{\sfr}{{\rm SFR}}

\newcommand{\ssfr}{{\rm sSFR}}
\newcommand{\lir}{L_{\rm IR}}

\newcommand{\leight}{L_8}
\newcommand{\ireight}{{\rm IR8}}

\newcommand{\msun}{{\rm M}_\odot}

\newcommand{\kpc}{{\rm kpc}}
\newcommand{\Gyr}{{\rm Gyr}}

\newcommand{\yr}{{\rm yr}}
\newcommand{\dex}{{\rm dex}}
\newcommand{\mstar}{M_\ast}

\newcommand{\tdust}{T_{\rm dust}}

\newcommand{\uvj}{$UVJ$\xspace}

\newcommand{\mean}[1]{\left<#1\right>}

\newcommand{\kelvin}{{\rm K}}
\newcommand{\zphot}{z_{\rm phot}}
\newcommand{\zspec}{z_{\rm spec}}

\defcitealias{schreiber2015}{S15}
\defcitealias{chary2001}{CE01}

\begin{document}

\title{The ALMA Redshift 4 Survey (AR4S). \\
I. The massive end of the $z=4$ main sequence of galaxies}
\titlerunning{AR4S: The massive end of the $z=4$ main sequence of galaxies}
\authorrunning{C.~Schreiber et al.}

\author{C.~Schreiber\inst{1,2}, M.~Pannella\inst{3,2}, R.~Leiton\inst{4,5}, D.~Elbaz\inst{2}, T.~Wang\inst{2,6}, K.~Okumura\inst{2} and I.~Labb\'e\inst{1}
}

\institute{
        Leiden Observatory, Leiden University, NL-2300 RA Leiden, The Netherlands\\
        \email{cschreib@strw.leidenuniv.nl}
        \and Laboratoire AIM-Paris-Saclay, CEA/DSM/Irfu - CNRS - Universit\'e Paris Diderot, CEA-Saclay, pt.~courrier 131, F-91191 Gif-sur-Yvette, France     \and Faculty of Physics, Ludwig-Maximilians Universit\"at, Scheinerstr.\ 1, 81679 Munich, Germany
        \and Instituto de F\'isica y Astronom\'ia, Universidad de Valpara\'iso, Avda.~Gran Breta\~na 1111, Valparaiso, Chile
        \and Astronomy Department, Universidad de Concepci\'on, Concepci\'on, Chile
        \and School of Astronomy and Space Sciences, Nanjing University, Nanjing, 210093, China
}

\date{Received 21 june 2016; accepted 16 December 2016}

\abstract {We introduce the ALMA Redshift 4 Survey (AR4S), a systematic ALMA survey of all the known galaxies with stellar mass ($\mstar$) larger than $5\times10^{10}\,\msun$ at $3.5<z<4.7$ in the GOODS--{\it south}, UDS and COSMOS CANDELS fields. The sample we have analyzed in this paper is composed of $96$ galaxies observed with ALMA at $890\,\um$ ($180\,\um$ rest-frame) with an on-source integration time of $1.3\,{\rm min}$ per galaxy. We detected $32\%$ of the sample at more than $3\sigma$ significance. Using the stacked ALMA and \herschel photometry, we derived an average dust temperature of $40\pm2\,\kelvin$ for the whole sample, and extrapolate the $\lir$ and $\sfr$ for all our galaxies based on their ALMA flux. We then used a forward modeling approach to estimate their intrinsic $\ssfr$ distribution, deconvolved of measurement errors and selection effects: we find a linear relation between $\sfr$ and $\mstar$, with a median $\ssfr=2.8\pm0.8\,\Gyr$ and a dispersion around that relation of $0.28\pm0.13\,\dex$. This latter value is consistent with that measured at lower redshifts, which is proof that the main sequence of star-forming galaxies was already in place at $z=4$, at least among massive galaxies. These new constraints on the properties of the main sequence are in good agreement with the latest predictions from numerical simulations, and suggest that the bulk of star formation in galaxies is driven by the same mechanism from $z=4$ to the present day, that is, over at least $90\%$ of the cosmic history. We also discuss the consequences of our results on the population of early quiescent galaxies. This paper is part of a series that will employ these new ALMA observations to explore the star formation and dust properties of the massive end of the $z=4$ galaxy population.
}


\maketitle

\section{Introduction}

It is now well established that galaxies have formed most of their stars around $z=2$ (e.g., \citealt{madau2014} and references therein) and that the majority of this star-formation activity happens in galaxies that belong to the main sequence (MS) of star-forming galaxies, a tight correlation between the galaxies' stellar mass ($\mstar$) and star-formation rate ($\sfr$) \citep[e.g.,][]{noeske2007,elbaz2007,daddi2007-a,pannella2009-a,elbaz2011,schreiber2015}. At a given stellar mass, the typical $\sfr$ of galaxies belonging to this sequence has evolved dramatically though time, continuously going down by about an order of magnitude from $z=2$ to the present day \citep[e.g.,][]{daddi2007-a}. This is can be explained by a progressive depletion of gas reservoirs \citep[e.g.,][]{daddi2008,tacconi2010} together with an additional decline of the star-formation efficiency over the same time period \citep[e.g.,][]{schreiber2016-b}.

The situation at $z>2$ is less clear. Over the past decade, most of our knowledge of the early Universe has been based on observations of the stellar emission in the rest-frame UV-to-optical, which allow detecting galaxies even beyond the re-ionization era \citep[e.g.,][]{stark2009,bouwens2012,salmon2015,oesch2016}. But in the absence of direct mid- or far-IR measurement, accurately correcting for absorption by interstellar dust is challenging. Because of the known correlation between mass and attenuation \citep[e.g.,][]{pannella2015}, this is particularly important if one wants to study the massive end of the galaxy population \citep[e.g.,][]{spitler2014}. While small in numbers, these massive galaxies (say, more massive than the Milky Way, with $\mstar > 5\times10^{10}\,\msun$) contribute about half of the star-formation activity in the Universe at any $z\leq3$ \citep[e.g.,][]{schreiber2015}.

To date, observations of distant galaxies in the far-IR or submillimeter with single dish instruments only allowed the detection of the brightest objects, that are experiencing extreme star-formation episodes and may not be representative of the overall population \citep[e.g.,][]{pope2006,capak2011,riechers2013}. Through stacking of carefully chosen samples, the average $\sfr$ and gas mass can be derived \citep[e.g.,][]{magdis2012,heinis2014,pannella2015,schreiber2015,bethermin2015-a,tomczak2016}, although these need to be corrected for the effect of source blending and clustering, which is not trivial. Stacking also allows determining the scatter of the stacked properties \citep{schreiber2015}, however this requires higher $S/N$ levels which cannot be reached beyond $z=3$ even with the deepest \herschel images.

Recently, the deployment of the Atacama Large Millimeter Array (ALMA) has allowed us to move forward and detect (or put stringent constraints on) the dust emission of less extreme galaxies at $z=4$ and beyond \citep[e.g.,][]{capak2015,maiolino2015,scoville2016}. However, these first efforts were mostly focused on Lyman break galaxies (LBGs) which are necessarily blue, less dust-obscured and found preferentially at the lowest masses \citep[e.g.,][]{spitler2014,wang2016}. At present, we are still lacking a complete census of the massive galaxy population at these redshifts, and this is the gap we intend to fill with this paper.

We therefore introduce here the ALMA Redshift 4 Survey (AR4S\footnote{Read ``aras'', which is the French word for macaw birds.}), a complete survey of massive galaxies in the {\it Hubble Space Telescope} (\hst) CANDELS fields at $z\sim4$ with the ALMA telescope. By probing the rest-frame $180\,\um$ emission, we can put direct constraints on the infrared luminosity, and therefore on the $\sfr$, of about a hundred galaxies at these epochs. We describe the sample in \rsec{SEC:sample} and the reduction of the ALMA data in \rsec{SEC:alma}. We stacked the ALMA fluxes and \herschel images in \rsec{SEC:sfr} to measure the average dust temperature in our sample, and use it to extrapolate the $\sfr$s for all our galaxies. We then discuss the location of our galaxies on the $\sfr$--$\mstar$ plane in \rsec{SEC:ms}, and model their $\ssfr$ distribution in \rsecs{SEC:model} and \ref{SEC:modelres} to provide the first robust constraints on the normalization and scatter of the MS at $z=4$. Lastly, we briefly discuss the existence of quiescent galaxies in our sample in \rsec{SEC:qgal}.

In the following, we assume a $\Lambda$CDM cosmology with $H_0 = 70\ {\rm km}\,{\rm s}^{-1} {\rm Mpc}^{-1}$, $\Omega_{\rm M} = 0.3$, $\Omega_\Lambda = 0.7$ and a \cite{salpeter1955} initial mass function (IMF), to derive both star-formation rates and stellar masses. All magnitudes are quoted in the AB system, such that $m_{\rm AB} = 23.9 - 2.5\log_{10}(S_{\!\nu}\ [\uJy])$.

\section{Sample selection \label{SEC:sample}}

\subsection{Observed sample}

We drew our sample from the CANDELS \hst $H$-band catalogs in the three fields covered by deep \herschel imaging and accessible with ALMA, namely GOODS--{\it south} \citep{guo2013-a}, UDS \citep{galametz2013} and COSMOS (Nayyeri et al.~in prep.). Photometric redshifts and stellar masses for all galaxies in this parent sample were derived in \cite{schreiber2015}. From there, our sample is selected purely on stellar mass ($\mstar > 5\times10^{10}\,\msun$) and photometric redshift ($3.5 < z < 4.7$) to ensure the highest completeness, resulting in $110$ galaxies. We did not try to segregate actively star-forming galaxies from quiescent ones, since the known selection techniques (e.g., the \uvj selection; \citealt{williams2009}) are uncertain at $z\geq4$ (see discussion in \rsec{SEC:qgal}). Instead, we targeted all galaxies regardless of their potential star-formation activity. We added to our target list $16$ lower mass galaxies with a spectroscopic redshift within $3.5 < z < 4.7$; these objects are part of a second sample and will not be discussed further in the present paper.

The median $H$-band magnitude of our sample is $25.2$, meaning that these galaxies are faint but still far above the $5\sigma$ point-source limiting magnitude of $H=27$. However, the $H$-band probes the rest-frame UV emission at $z=4$. Even though the CANDELS \hst images are the deepest available to date, this implies that our parent sample is very likely biased against the most heavily obscured galaxies at these redshifts. Indeed, a population of $H$-dropouts (but \spitzer IRAC bright) galaxies has recently been identified in these fields \citep{wang2016}. If indeed at $z\sim4$, these would represent up to $20\%$ of our selected sample and would therefore only impact our results marginally. These galaxies are currently being observed as part of another ALMA program, and will be discussed in a future work (Wang et al.~in prep.). We estimate a similar completeness based on the samples of ultra-red galaxies missed by \hst from \cite{huang2011} and \cite{caputi2012}.

\subsection{Cleaning the sample}

With these short integration times, we chose to include in our selection all the potential $z=4$ massive galaxies, regardless of the quality of their photometry. While most galaxies in the sample do have clean measurements in all bands, we identify $14$ likely spurious or contaminated sources. Ten sources, the majority, have their \spitzer IRAC photometry clearly contaminated by a bright neighboring source (either a star or an extended nearby galaxy). Their redshifts and stellar masses are unreliable, and they are consistently not detected on the ALMA images. Two sources are each very close (<$1\arcsec$) to another nearby galaxy which is also part of our $z=4$ sample. Because their redshifts are consistent with being identical with that of their neighbor, and because the ALMA emission originates from the barycenter of the two sources, we choose to consider them as a single object and re-measure their UV-NIR fluxes and stellar mass in a larger aperture, as described in the next section. Finally, one source is out the \spitzer IRAC coverage, and another is partially truncated at the border of the \hst $H$-band image.

Excluding these objects, we ended up with a final sample of $96$ good quality $z=4$ massive galaxies, which we analyzed in the following.

\subsection{Robustness of the redshift and masses determinations \label{SEC:zm}}

Since our targets are relatively faint, and because the $H$ band is tracing the rest-frame UV, it is reasonable to wonder if the template fitting approach used by CANDELS to extract the photometry \citep[using TFIT,][]{laidler2007} is adequate, and not missing part of the flux. For example, the well-known $z=4$ starburst GN20 has a patchy dust geometry \citep{hodge2012}, and its rest-UV emission is not representative of the true extent of the galaxy. As a check, we therefore re-measured the photometry of all our targets, specifically looking for missing flux outside of the CANDELS $H$-band segmentation. We summed the flux on the UV-to-NIR images in large apertures of $0.9\arcsec$ or more, depending on the apparent morphology of the galaxy in the various bands, and indeed find on average an additional $30\%$ flux in the bluest bands. Since the galaxies are not resolved on the \spitzer IRAC images, the IRAC fluxes (probing the rest-optical and constraining the stellar mass) are essentially unaffected. We re-analyzed this new photometry as in \cite{schreiber2015} to obtain redshifts and stellar masses: reassuringly, the photometric redshifts are not affected in any significant way ($10\%$ scatter in $\Delta z/(1+z)$), and stellar masses are also globally unchanged ($0.06\,\dex$ increase on average, with $0.18\,\dex$ scatter). The scatter in these quantities is comparable to the one we observe when cross-matching our sample to the 3DHST catalogs \citep{skelton2014}: $9\%$ for the redshifts and $0.2\,\dex$ for the masses.

In the following, to avoid adding extra noise to our selection, we therefore continued to use the original CANDELS photometry except for  four galaxies which were clearly missing a large fraction of their flux because of overly aggressive deblending (see also previous section). These are \texttt{ID}$=$$23751$ in GOODS--{\it south}, $5128$ and $35579$ in UDS, and $27853$ in COSMOS.

\section{ALMA observations and reduction \label{SEC:alma}}

\subsection{Data}

\begin{figure}
\begin{center}
\includegraphics[width=9cm]{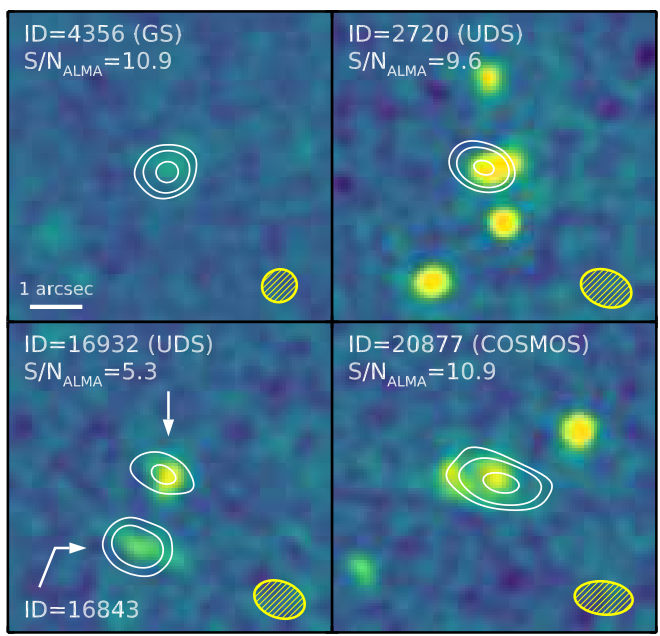}
\end{center}
\caption{Examples of ALMA-detected galaxies in our sample. The \hst F160W image is shown in the background, smoothed by a $0.3\arcsec$ FWHM Gaussian to reveal extended features. The white contours show the extent of the ALMA emission ($3$, $5$, $10$ and $17\,\sigma$) after applying tapering. The size and orientation of the ALMA clean beam is given with a yellow hatched region (FWHM). The CANDELS ID of each target is given at the top of each cutout, together with the $S/N$ of the integrated ALMA flux. \label{FIG:cutouts}}
\end{figure}

Our targets were observed with dedicated ALMA pointings in band 7, with a central wavelength of $\sim$$890\,\um$ ($888.1\,\um$, $337.8\,{\rm GHz}$), integrating $1.3$ minutes on each galaxy for a total observing time of $6$ hours. We reduced the data into calibrated visibilities using the CASA pipeline (version 4.3.1), and produce cleaned continuum images for visual inspection. The achieved angular resolution varies from one field to the other, ranging from $0.3$ to $0.7\,\arcsec$ (FWHM of the minor axis of the beam), that is, $2$ to $5\,\kpc$ at $z=4$. To measure the effective noise level on comparable grounds, we tapered the longest baselines to reach an homogeneous angular resolution of $0.7\,\arcsec$ and find an RMS of $0.15$ to $0.22\,\mJy$. Examples of clear detections from the tapered images are given in \rfig{FIG:cutouts}.

\subsection{Flux measurements}

We measured the $890\,\um$ flux of all our targets directly in the $(u,v)$ plane using the \texttt{uvmodelfit} procedure from the CASA pipeline. The sources were first modeled with an elliptical Gaussian profile of variable total flux, centroid, width, axis ratio and position angle. When the $S/N$ is too low, the fit becomes unstable and tends to return large position offsets ($>2\arcsec$). In these cases, we discarded the fit and used a simpler model where the position is frozen to that of the HST counterpart, and the size is fixed to the median size of the high $S/N$ galaxies. In both cases the adopted flux uncertainty is the formal uncertainty returned by \texttt{uvmodelfit}, and the fluxes and uncertainties are corrected for primary beam attenuation a posteriori.

The fluxes measured with this method are on average $37\%$ higher than the peak fluxes read from the cleaned images (with tapering), indicating that most of our targets are resolved by ALMA (the median half-light radius returned by \texttt{uvmodelfit} is $0.3\arcsec$, or $2\,\kpc$ at $z=4$). Consequently, the uncertainties are also larger than the image RMS, with an average of $0.31\,\mJy$. This corresponds to a $3\sigma$ detection limit of $\sfr\simeq294\,\msun/\yr$ at $z=4$ (see \rsec{SEC:sfr} for the conversion to $\sfr$).

We checked the accuracy of these measurements by fitting fake sources at random positions in the field of view, devoid of significant detection, and find a mean flux of $0.02\pm0.02\,\mJy$, indicating that our fluxes are not biased. The RMS of these fake measurements is $0.3\,\mJy$, and is therefore fully consistent with our average flux uncertainty. As a cross-check, we also compared our flux determination for the galaxy \texttt{ID=23751} against the published value from the ALESS program \citep{hodge2013}. These two independent measurements agree on a total flux of $8.0\,\mJy$ with uncertainties of $0.4$ and $0.6\,\mJy$ in our catalog and that of ALESS, respectively. More details on the reduction and flux measurements will be provided together with the complete catalog in a forthcoming paper (Leiton et al.~in prep.).

\subsection{Properties of the detected and non-detected galaxies}

Of all targets in the present sample, we detect $46$, $30$ and $17$ galaxies at $>$$2$, $3$ and $5\sigma$ significance, respectively. The mean flux of the sample is $1.00 \pm 0.04\,\mJy$ including non-detections.

Both detected and non-detected galaxies have a similar redshift distribution; a Kolmogorov-Smirnov (KS) test gives a probability $>$99\% that they share the same photometric redshift distribution. The average redshift is $\mean{z}=3.99$, with a standard deviation of $0.36$. The detections tend to have slightly fainter $H$-band magnitudes, although this is barely significant (the KS probability of same distribution is still $30\%$). However the ALMA detections have significantly higher stellar masses than the non-detections: ALMA-detected galaxies constitute the majority of galaxies at $\log\mstar > 11.3$, while non-detections become dominant at $\log\mstar < 10.9$ (KS test $<$1\%). This is most likely a consequence of the $\sfr$--$\mstar$ correlation, as we will show in the following.

\subsection{Note on the astrometry}

In their survey of the \hubble Ultra Deep Field (HUDF, in the center of GOODS--{\it south}), \cite{dunlop2016} reported a systematic position offset of $\Delta\delta = \delta_{\rm HST} - \delta_{ALMA} = +0.24\arcsec$ between the ALMA detections and their \hst counterparts, with the ALMA emission observed south of the \hst. \cite{rujopakarn2016} confirmed this offset using the JVLA and 2MASS, measuring $\Delta\delta = +0.26\pm0.13\arcsec$, and suggesting an issue in the absolute astrometry of the \hst images. None of our targets falls in the HUDF, so we cannot directly double check their result. However, selecting our $3\sigma$ detections in the whole GOODS field, we consistently find a median systematic shift of $\Delta\delta=+0.15\pm0.05\arcsec$ in GOODS--{\it south}. In addition, we find a shift of $\Delta\alpha=-0.14\pm0.05\arcsec$ in UDS, and no significant shift in COSMOS. These shifts are relatively small compared to the ALMA beam, and vary substantially from one source to another ($0.2\arcsec$ scatter for the $3\sigma$ detections). We therefore do not attempt to correct for it; since we fit for the centroid for most of our targets, this would have a negligible impact anyway.

\section{Results \label{SEC:results}}

\subsection{Dust temperature and star-formation rates \label{SEC:sfr}}

\begin{figure}
\begin{center}
\includegraphics[width=9cm]{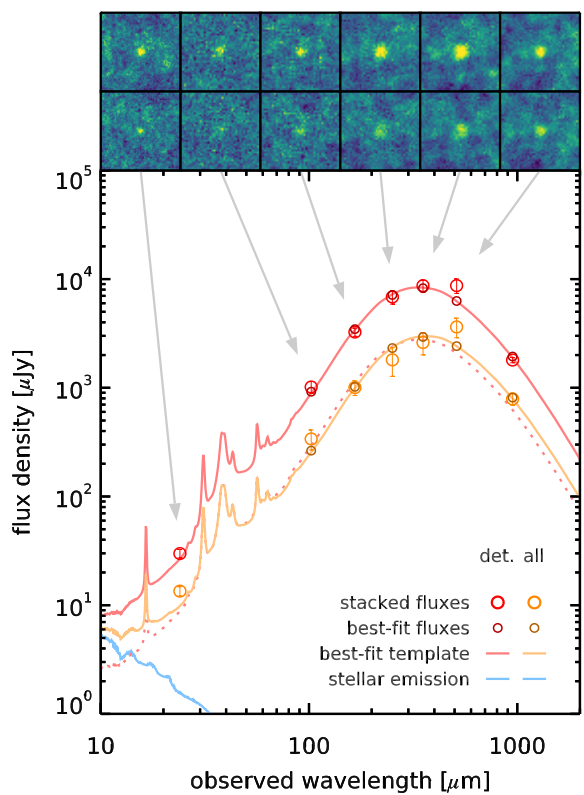}
\end{center}
\caption{Stacked mid- to far-infrared SED of our full sample (orange) and only the $3\,\sigma$ ALMA detections (red). The large open circles with error bars are the stacked fluxes, with SPIRE fluxes corrected for flux boosting from clustering (see text). Darker and smaller circles show the best-fit model fluxes, and the corresponding template is shown with a solid line in the background (NB: MIPS $24\,\um$ was not used in the fit). The blue solid line is the stellar emission, estimated by fitting the stacked UV-to-NIR photometry with FAST \citep{kriek2009}. The dotted line is the best-fit SED of the ALMA detections, rescaled to the $\lir$ of the whole sample for easier comparison. The stacked cutouts are shown at the top of the plot, both for the the ALMA detections (top) and the full sample (bottom). The arrows help identify which flux measurement they correspond to. \label{FIG:sed}}
\end{figure}

To determine the $\sfr$ of the galaxies in our sample, we need to extrapolate the $8$-to-$1000\,\um$ infrared luminosity ($\lir$) of all galaxies from the $890\,\um$ flux alone. Indeed, most of the galaxies of our sample are undetected in other MIR or FIR bands, and their IR SED cannot be constrained individually. However, we can stack them together on the \spitzer and \herschel images to recover their average SED, which can then be used to infer $\lir$ assuming it is representative of the whole sample. As in \cite{schreiber2015}, we median stack the images and measure the flux by fitting a point spread function (PSF) model at the center of the stacked image, with a freely varying background level. Because of the poor angular resolution of the images, SPIRE fluxes are boosted by the contribution of clustered nearby objects, and we statistically correct for this boost following \cite{schreiber2015}. Uncertainties are determined by bootstrapping the sample.

We show in \rfig{FIG:sed} the median-stacked SED from \spitzer, \herschel and ALMA, separately for the full sample ($96$ galaxies) and for the $3\sigma$ ALMA detections only ($30$ galaxies). The stacked signal is stronger when we only consider the ALMA detections, and the corresponding stacked fluxes form a coherent SED in all bands. However, this sample is probably not representative of the whole population: the dust temperature ($\tdust$) is known to increase for brighter galaxies above the MS \citep{magnelli2014}, which are preferentially selected in such flux-limited samples \citep[e.g.,][]{rodighiero2011}. Therefore, we expect the sub-sample of ALMA-detected galaxies to have on average higher $\tdust$ than the rest of the population. That being said, the stacked SED of the full sample is more uncertain but does not appear to differ significantly from that of the ALMA detections.

To quantify this, we use the dust SEDs introduced in \cite{schreiber2016-b} and fit the stacked photometry. This library has three free parameters: $\lir$, $\tdust$ and $\ireight\equiv\lir/\leight$. Since we have no data to constrain $\leight$ (the rest-frame $8\,\um$ luminosity), we keep this last parameter fixed at $\ireight=8$ (\citealt{reddy2012}; Schreiber et al.~in prep.) and only fit for $\lir$ and $\tdust$. We find that this model provides a good description of our data, with a reduced $\chi^2$ of $1.3$ for both samples. The best-fit dust temperatures are $\tdust=40\pm2$ and $43\pm1\,\kelvin$ for the full sample and for the ALMA detections, respectively. Therefore we do find that our ALMA detections have on average higher $\tdust$ values, although the difference is only $1.3\sigma$ and mild: it would result in a difference of only $0.1\,\dex$ when extrapolating $\lir$ from the $890\,\um$ flux. Consequently, we choose here to assume a single dust temperature of $\tdust=40\,\kelvin$ for all galaxies, and expect our individual $\sfr$s to be systematically biased by up to $0.1\,\dex$. We also note that high dust temperatures around $40\,\kelvin$ have already been reported in the recent literature for $z=4$ galaxies \citep[e.g.,][]{dacunha2015,bethermin2015-a}.

Although the stacked \herschel fluxes were corrected for flux boosting by clustering, we find that the $500\,\um$ fluxes are systematically above our best-fit SED. This suggests that our correction, which was derived as an average correction from lower-redshift samples, is insufficient at least in this band. To check that our conclusions are not affected by this potential bias, we re-run the fit excluding the SPIRE bands and find higher $\tdust$ values by only $0.6\,\kelvin$, which is well within our error bars. We therefore conclude that our $40\,\kelvin$ SED is not significantly biased by clustering.

Using this SED, we extrapolated the $\lir$ for all our targets and convert it into $\sfr$ with the \cite{kennicutt1998-a} conversion factor. We also include the contribution of the non-obscured UV light, although it is always negligible.

\subsection{The \texorpdfstring{$z=4$}{z=4} main sequence \label{SEC:ms}}

\begin{figure*}
\includegraphics[width=9cm]{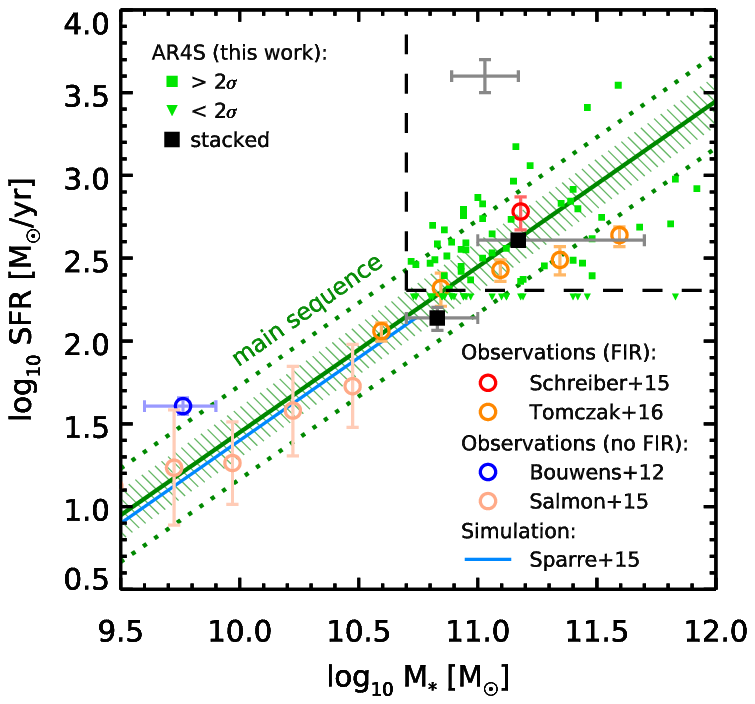}
\includegraphics[width=9cm]{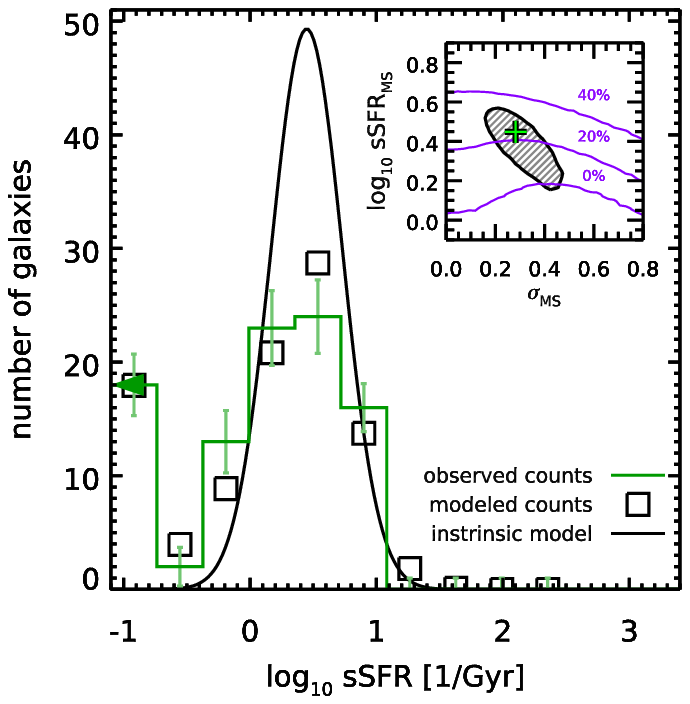}
\caption{{\bf Left:} Relation between the star-formation rate ($\sfr$) and stellar mass ($\mstar$) at $z=4$. Our ALMA sample is shown with small green squares ($>$$2\sigma$), and downward-facing triangles ($<$$2\sigma$). The gray cross at the top of the plot indicates the systematic uncertainty affecting our measurements (see text). The median $\sfr$ of our sample in two mass bins is shown with black squares (correcting the median for the contaminant population, see text). The green lines are our estimate of the $z=4$ MS locus (solid line) and scatter (dotted lines) as determined from modeling the $\ssfr$ distribution (see text and figure on the right), and the hatched area in the background indicates the uncertainty on the MS locus. Observations from the literature are shown with open circles, from \cite{bouwens2012} (blue) and \cite{salmon2015} (pink) who both derived their $\sfr$s without MIR or FIR observations, while \cite{tomczak2016} (orange, $z=3.5$) and \cite{schreiber2015} (red) both used \herschel stacking. The latter are corrected by $-0.1\,\dex$ to account for the difference between the mean and median. Finally, the MS relation found in the Illustris simulation is shown as a blue line \citep{sparre2015}. {\bf Right:} Distribution of $\ssfr = \sfr/\mstar$ for our sample. The observed counts (including non-detections) are shown with a green histogram and error bars. The best-fit modeled intrinsic distribution is shown in black in the background, and the modeled counts (simulating measurement uncertainties and selection effects) are shown with open squares. The $1\sigma$ confidence region for the model parameters is shown in inset, where the green cross gives the position of minimum $\chi^2$, and the purple lines indicate the contaminant fraction required by the fit (see text). \label{FIG:sfrms}}
\end{figure*}

We show the location of our galaxies on the $z=4$ $\sfr$--$\mstar$ plane in \rfig{FIG:sfrms} (left). At first order, the detections appear to be distributed around the expected location of the MS, as inferred from previous \herschel stacking \citep[e.g.,][]{heinis2014,schreiber2015,tomczak2016}, with a tendency for higher mass galaxies to have overall higher $\sfr$s. However, because of our relatively low detection rate, deriving conclusions on the locus and scatter of the MS from this figure alone can prove difficult.

Instead, we use a forward modeling approach where we model the observed distribution of $\ssfr = \sfr/\mstar$ (shown on the right panel of the same figure) for both detected and non-detected galaxies, without stacking them. The model assumes an intrinsic functional form and includes all quantifiable sources of selection and observational effects (see, e.g., \citealt{mullaney2015}). We describe the method we employ in the next section.

\subsection{Modeling the \texorpdfstring{$\ssfr$}{sSFR} distribution \label{SEC:model}}

In the following, we will make the distinction between ``intrinsic'' and ``observed'' mock quantities. The former are the real galaxy properties, free of measurement error and systematics, while the latter are an attempt at simulating the real observations we report in this work, after including all sources of noise. Owing to the modest size of our sample, we settle for a simple three-parameter model where the $\ssfr$ of star-forming galaxies is distributed according to a log-normal law of logarithmic width $\sigma_{\rm MS}$, and median $\ssfr_{\rm MS}$ \citep[e.g.,][]{rodighiero2011,sargent2012,schreiber2015,ilbert2015}. In addition, we allow a fraction $f_{\rm c}$ of galaxies in the model to be flagged as ``contaminants'', in which case their $\ssfr$ is set to zero. This last step takes into account two sources of contamination in our sample. First and foremost is the existence of quiescent galaxies, which have no star formation by definition, and which could contribute as much as $30\%$ of our sample \citep{straatman2014,spitler2014}. A secondary source of contamination are redshift interlopers and brown dwarfs which we would have failed to identify with the available photometry, and which are expected to be substantially fainter at submillimeter wavelengths.

We created a grid for the first two parameters (as displayed in the inset of \rfig{FIG:sfrms}, right), and model the observed $\ssfr$ distribution for each cell of that grid. To do so, we generated a mock catalog of $100\,000$ galaxies to which we attribute an observed redshift and stellar mass by drawing randomly and independently from their respective distributions in our real $z=4$ sample. We then perturbed these redshifts and masses with a Gaussian distribution of width $\Delta z/(1+z) = 6.5\%$ and $\Delta \log_{10}{\mstar/\msun} = 0.14\,\dex$, respectively, to obtain the associated intrinsic quantities ($1/\!\sqrt{2}$ times the values we obtained in \rsec{SEC:zm}). It is important to take into account these sources of uncertainty to constrain the intrinsic scatter of the main sequence; if we had not modeled them, our measured intrinsic scatter (see next section) would have been $0.05\,\dex$ higher. Conversely, if we had used more pessimistic values of $\Delta z/(1+z) = 10\%$ and $\Delta \log_{10}{\mstar/\msun} = 0.2\,\dex$, our measured scatter would have been smaller by $0.02\,\dex$. It turns out this variation is well within the statistical uncertainties, so our conclusions do not strongly depend on these values.

Given these intrinsic redshifts and masses, we then use the assumed MS model to derive the intrinsic $\ssfr$ of each mock galaxy, including a mild redshift evolution of $\mean{\ssfr} \sim (1+z)^{1.5}$ \citep{schreiber2015}. The next step is to infer the $890\,\um$ flux. To do so, we first converted the $\ssfr$ into $\lir$ using the intrinsic stellar mass and the \cite{kennicutt1998-a} conversion factor. We then attributed an intrinsic $\tdust$ to each mock galaxy, with an average of $40\,\kelvin$ and a Gaussian scatter of $3\,\kelvin$ (as observed at lower redshifts, e.g., \citealt{ciesla2014}), and used the corresponding SED to infer the intrinsic ALMA flux. The observed flux is obtained simply by adding a random Gaussian noise, whose amplitude is drawn from the flux uncertainty distribution. Given these mock observed fluxes, we applied the same procedure as in the previous section to derive the observed $\sfr$ (including, in particular, the assumption of a fixed $\tdust = 40\,\kelvin$), and finally obtained the observed $\ssfr$ by dividing this value by the observed $\mstar$. We applied the same procedure to a second mock catalog where the intrinsic ALMA fluxes are all set to zero to model the contaminant population, and build histograms of the observed $\ssfr$ for both mock catalogs. These histograms are normalized to the total number of simulated galaxies ($100\,000$).
A linear combination of these two histograms is then fit to our observed data, adjusting their respective normalization to match the total number of observed galaxies, and the relative weight of each histogram allows us to derive $f_{\rm c}$. In this fit, the uncertainties on the $\ssfr$ histograms were derived by repeatedly perturbing all the $\ssfr$s by their respective uncertainties, and computing the standard deviation of the counts in each bin. As a last step we noted down the $\chi^2$ of the fit and proceeded to the next cell in the grid.

\subsection{Result \label{SEC:modelres}}

In \rfig{FIG:sfrms} (right) we show the intrinsic $\ssfr$ distribution that produces the smallest $\chi^2$ when compared to our observations. This distribution has $\sigma_{\rm MS} = 0.28^{+0.14}_{-0.12}\,\dex$ and $\log_{10}(\ssfr_{\rm MS}\,[1/\Gyr]) = 0.45^{+0.11}_{-0.18}$, and the fit required a contaminant fraction of $f_{\rm c} = 25\%$.

This median $\ssfr$ is lower but still consistent with our previous determination from \herschel stacking of $\log_{10}(\ssfr_{\rm MS}\,[1/\Gyr])=0.6\pm0.1$ \citep{schreiber2015}, and the width of the distribution appears unchanged from the value of $\sigma_{\rm MS} = 0.31 \pm 0.02\,\dex$ we obtained at $z\sim1$ (see also \citealt{ilbert2015,guo2015}). Overall, this confirms the theoretical expectation that the MS paradigm still holds in the early Universe \citep[e.g.,][]{sparre2015}. This agreement is both qualitative and quantitative, since our determination of the locus and scatter of the MS is also in agreement with the prediction of the latest numerical simulations. \cite{sparre2015} indeed found a scatter of $0.2$--$0.3\,\dex$ at $z=4$ in the Illustris simulation with a scatter closer to $0.3\,\dex$ at high mass, although at these redshifts they lack the volume to constrain the $\mstar > 3\times10^{10}\,\msun$ population. Similarly, \cite{mitchell2014} found a $0.27\,\dex$ scatter in the GALFORM semi-analytic model at $z=3$, this time with an increase toward higher redshifts (contrary to Sparre et al.~who found, if anything, an opposite evolution). Our data are still insufficiently deep to rule out current models based on the scatter only, however the $z=3$ MS normalization of Mitchell et al.~(as also discussed in their work) is particularly low. Assuming a typical redshift dependence of $\ssfr \sim (1+z)^{2.5}$ \citep[e.g.,][]{dekel2013}, it is inconsistent with our observations at the $3.5\,\sigma$ level.

As a check, we then compute the median $\sfr$ of our sample in two bins of stellar mass above and below $\mstar = 10^{11}\,\msun$ (still including non-detections), and divide the obtained values by $(1-f_{\rm c})$ to statistically remove the contribution of the contaminants. The obtained $\sfr$s, $138\pm22$ and $406\pm25\,\msun/\yr$, are reported on \rfig{FIG:sfrms} (left), and clearly illustrate the positive correlation between $\sfr$ and $\mstar$. The low-mass point is found below the $\sfr$--$\mstar$ relation derived from the $\ssfr$ modeling, although at a significance of only $0.7\sigma$. This may suggest that the fraction of contaminant is higher among low-mass galaxies, which can be expected given that these are about one magnitude fainter and are therefore more likely to be wrongly characterized.

\subsection{Contaminants: quiescent galaxies or redshift interlopers? \label{SEC:qgal}}

The contaminant fraction $f_{\rm c}=25\%$ is low, and remains consistent with the observed number of quiescent galaxies at these redshifts \citep{muzzin2013,straatman2014}. We cannot rule out, however, that a substantial fraction of these ``contaminants'' could be low redshift interlopers. At first order, we can obtain an estimate of how many such outliers contaminate our sample by comparing our photometric redshifts ($\zphot$) at $3.5<\zphot<5$ against spectroscopic determinations ($\zspec$) from the literature. Since only one of the AR4S galaxy has a $\zspec=3.582$, we widen our selection to include $110$ galaxies of lower stellar masses. Among these, $4\%$ have $\zspec < 2$. This spectroscopic sample has substantially bluer colors than the typical AR4S galaxy (observed $(r-H) \sim 1$ and $3$, respectively), so we expect this fraction to be larger in AR4S, possibly up to $10\%$. Subtracting these outliers from our contaminant population would imply that quiescent galaxies contribute only $15\%$ of our sample, which would be more consistent with the extrapolation of the quiescent fraction from lower redshifts. On the other hand, as shown in \rfig{FIG:sfrms} (right, inset), $f_{\rm c}$ can range from $0$ to $40\%$ within the $1\sigma$ confidence interval for our two model parameters. We therefore refrain from drawing strong conclusions out of this number.

Independent constraints on the quiescent population could be provided by their rest-optical colors. However, as mentioned in \rsec{SEC:sample}, the standard \uvj selection of quiescent galaxies is uncertain at $z=4$ because the rest-frame $J$ ($1.2\,\um$) band shifts into the \spitzer IRAC $5.8\,\um$ band. While the $5.8\,\um$ observations are deep enough in the GOODS--{\it south} field, they are shallower in UDS and COSMOS, and because they require cryogenic cooling (which is no longer available on-board \spitzer) their depth has not improved during the past years. Probably as a consequence of this poorly constrained $V-J$ color, $5$ out of $30$ of our ALMA detections are classified as \uvj-quiescent (one of them being in GOODS--{\it south}).  Encouragingly, the fraction of \uvj quiescent galaxies is larger among the ALMA non-detections ($31$ out of $66$), but their stacked ALMA flux of $0.4\pm0.1\,\mJy$ suggests that not all are truly quiescent.

The red colors of these galaxies could be mistaken for a quiescent stellar population if these galaxies contain dust-obscured AGNs \citep[e.g.,][]{donley2012}. Among the $5$ \uvj-quiescent ALMA detections, only one of them is strongly detected in MIPS $24\,\um$, and none is detected in the available X-ray or radio images. Among the $31$ non-detections, we find one $24\,\um$ detection, one X-ray detection and one X-ray+radio detection. Therefore the majority of these galaxies do not show sign of AGN activity.

We conclude that the \uvj classification in CANDELS UDS and COSMOS is indeed unreliable at these redshifts, and until the launch of the {\it James Webb Space Telescope}, the only way to robustly identify these $z\geq4$ quiescent objects is to constrain their individual star-formation activity, either with deep NIR spectroscopy or ALMA high-frequency imaging.

\section{Conclusions}

We introduce AR4S, a systematic ALMA survey of all the known massive galaxies ($\mstar > 5\times10^{10}\,\msun$) at $z=4$ in the deepest fields observed with \hubble and \herschel. With only $6$ hours of telescope time, we detect $30$ out of $95$ galaxies: our strategy of targeting mass-selected samples is an order of magnitude more efficient at detecting high redshift galaxies compared to contiguous surveys \citep[e.g.,][]{dunlop2016,walter2016}. Detailed properties of the detected and non-detected galaxies, as well as the full catalog, will be discussed in a forthcoming paper (Leiton et al.~in prep.).

Using these new data, we first build the average dust SED of our sample and find an average dust temperature of $40\kelvin$, in agreement with recent studies of $z\sim4$ galaxies.

From this SED, we extrapolate the total $\lir$ (hence $\sfr$) from the ALMA fluxes of our individual objects, and model the observed $\ssfr$ distribution with a three-parameter model inspired by observations at lower redshifts. This analysis suggests that galaxies at $z=4$ follow a linear relation between $\sfr$ and $\mstar$, with a dispersion around that relation of $0.3\,\dex$. This value is the same as that measured at lower redshifts, which is proof that the MS is already in place at $z=4$, at least among massive galaxies. These new constraints on the properties of the MS are also in good agreement with the latest prediction from numerical simulations (e.g., Illustris).

Finally, our results are compatible with the existence of a population of massive quiescent galaxies early in the history of the Universe, although spectroscopic confirmation and deeper ALMA imaging would be required to draw reliable conclusions.
Future works with this data will include a study of the dust attenuation properties of the galaxies in our sample (Pannella et al.~in prep.) and the geometry of their star-forming regions (Elbaz et al.~in prep.).
\begin{acknowledgements}

Most of the analysis for this paper was done using {\tt phy++}, a free and open source C++ library for fast and robust numerical astrophysics (\hlink{cschreib.github.io/phypp/}).

This paper makes use of the following ALMA data: ADS/JAO.ALMA\#2013.1.01292.S. ALMA is a partnership of ESO (representing its member states), NSF (USA) and NINS (Japan), together with NRC (Canada) and NSC and ASIAA (Taiwan), in cooperation with the Republic of Chile. The Joint ALMA Observatory is operated by ESO, AUI/NRAO and NAOJ.

This work is based on observations taken by the CANDELS Multi-Cycle Treasury Program with the NASA/ESA \hst, which is operated by the Association of Universities for Research in Astronomy, Inc., under NASA contract NAS5-26555.

R.L.~acknowledges the financial support from FONDECYT through grant 3130558.

T.W.~acknowledges the financial support from the European Union Seventh Framework Program (FP7/2007-2013) under grant agreement No.~312725 (ASTRODEEP).
\end{acknowledgements}

\bibliographystyle{aa}
\bibliography{../bbib/full}

\end{document}